\documentclass[aps,graphicx,twocolumn]{revtex4}
\usepackage{graphicx}
\usepackage{color}

\begin{document}

\title{Long-distance entanglement purification for quantum communication}

\author{Xiao-Min Hu,$^{1,4}$ Cen-Xiao Huang,$^{1,4}$ Yu-Bo Sheng,$^{2,3,5}$\footnote{shengyb@njupt.edu.cn} Lan Zhou,$^{2,3,5}$ Bi-Heng Liu,$^{1,4}$\footnote{bhliu@ustc.edu.cn} Yu Guo,$^{1,4}$ Chao Zhang,$^{1,4}$ Wen-Bo Xing,$^{1,4}$ Yun-Feng Huang,$^{1,4}$ Chuan-Feng Li,$^{1,4}$\footnote{cfli@ustc.edu.cn} Guang-Can Guo$^{1,2,4,5}$ }
\address{$^1$CAS Key Laboratory of Quantum Information, University of Science and Technology of China, Hefei, 230026, People's Republic of China\\
$^2$Institute of Quantum Information and Technology, Nanjing University of Posts and Telecommunications, Nanjing, 210003, People's Republic of China\\
$^3$College of Mathematics \& Physics, Nanjing University of Posts and Telecommunications, Nanjing, 210003, People's Republic of China\\
$^4$CAS Center For Excellence in Quantum Information and Quantum Physics, University of Science and Technology of China, Hefei, 230026, People's Republic of China\\
$^5$Key Lab of Broadband Wireless Communication and Sensor Network Technology, Nanjing University of Posts and Telecommunications, Ministry of Education, Nanjing, 210003, People's Republic of China\\}
\date{\today }
\begin{abstract}
High quality long-distance entanglement is essential for both quantum communication and scalable quantum networks. Entanglement purification is to distill high-quality entanglement from low-quality entanglement in a noisy environment and it plays a key role in quantum repeaters. The previous significant entanglement purification experiments require two pairs of low-quality entangled states and were demonstrated in table-top. Here we propose and report a high-efficiency and long-distance entanglement purification using only one pair of hyperentangled state. We also demonstrate its practical application in entanglement-based quantum key distribution (QKD). One pair of polarization spatial-mode hyperentanglement was distributed over 11~km multicore fiber (noisy channel). After purification, the fidelity of polarization entanglement arises from 0.771 to 0.887 and the effective key rate in entanglement-based QKD increases from 0 to 0.332. The values of Clauser-Horne-Shimony-Holt (CHSH) inequality of polarization entanglement arises from 1.829 to 2.128. Moreover, by using one pair of hyperentanglement and deterministic controlled-NOT gate, the total purification efficiency can be estimated as $6.6\times 10^{3}$ times than the experiment using two pairs of entangled states with spontaneous parametric down conversion (SPDC) sources. Our results offer the potential to be implemented as part of a full quantum repeater and large scale quantum network.
\end{abstract}

\maketitle

Quantum entanglement~\cite{Horodecki2009} plays an essential role in both quantum communication~\cite{teleportation,Guo2019,QKD,QSDC} and scalable quantum networks~\cite{network}. However, the unavoidable environmental noise degrades entanglement quality. Entanglement purification is a powerful tool to distill high-quality entanglement from low-quality entanglement ensembles~\cite{purification1,purification2} and is the heart of quantum repeaters~\cite{repeater1}. Several significant entanglement purification experiments using photons~\cite{experiment1,experiment4}, atoms~\cite{experiment2}, and electron-spin qubits~\cite{experiment3} were reported. These experiments were all table-top and did not distribute entanglement over a long distance. Moreover, these experiments based on Ref.~\cite{purification1} were low efficiency for they require two copies of low-quality entangled states and consume at least one pair of low-quality entangled states even if the purification is successful. In optical systems, a spontaneous parametric down conversion (SPDC) source is commonly used to generate entangled states. The probabilistic nature of SPDC makes it still challenging to generate two clean pairs of entangled states simultaneously because of double-pair emission noise \cite{experiment4}.

Hyperentanglement~\cite{hyper1}, simultaneous entanglement with more than one degree of freedom (DOF), is more powerful and can be used to increase the channel capacity~\cite{highdimension1,highdimension2}. Hyperentanglement also fulfills quantum teleportation of a single photon with multiple DOFs~\cite{highdimension4,multiteleportation,Graham15}. The distribution of hyperentanglement were also reported~\cite{highdimension6,highdimension7}. Some entanglement purification protocols (EPPs) assisted by spatial mode DOF have been proposed~\cite{hyperpurification1,hyperpurification2,hyperpurification3}. Such deterministic entanglement purification usually requires the spatial or other entanglement to be more robust. The fidelity of purified polarization entanglement equals the fidelity of spatial entanglement, and this is essentially the transformation from spatial entanglement to polarization entanglement.

\begin{figure}[!tpbh]
\includegraphics[width=8cm,angle=0]{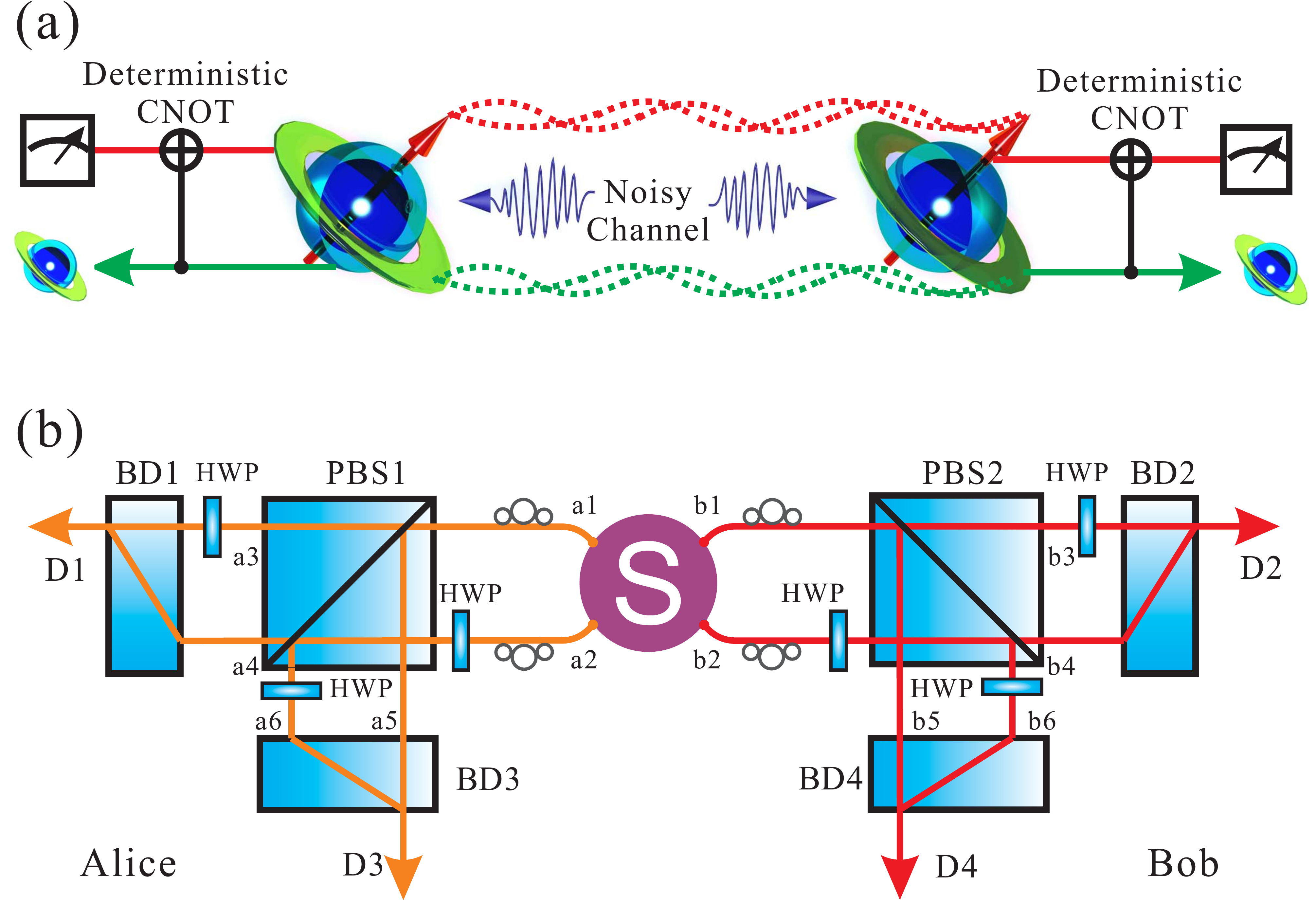}
\caption{\textbf{Protocol of entanglement purification using hyperentanglement.} \textbf{a}, General protocol of long-distance entanglement purification using hyperentanglement. After long-distance distribution, deterministic controlled-NOT (CNOT) operations are performed between different degrees of freedom of single particles. Then the target qubits (encoded in one degree of freedom) are measured. If we only keep the same measurement results, then the entanglement encoded in the other degree of freedom (control qubits) is purified. \textbf{b}, Purification protocol based on polarization-spatial mode hyperentanglement. The hyperentangled state $|\Phi^{+}\rangle_{ab}\otimes|\phi^{+}\rangle_{ab}$ is first generated by the entanglement source (S) and then distributed to Alice and Bob in the channel ($a_{1}b_{1}$, $a_{2}b_{2}$). After suffering from the channel noise, the entanglement purification is performed. The entanglement purification operation uses an HWP and a PBS. This essentially acts as a CNOT gate with a success probability of 100$\%$ between the polarization (target qubits) and spatial mode (control qubits) degree of freedom. After PBS1 and PBS2, we convert the spatial mode to polarization. Thus, we can verify its success by selecting the cases in which the two photons are in the output mode state D$_{1}$D$_{2}$ or D$_{3}$D$_{4}$. PBS--polarizing beam splitter, which transmits the H-polarized photon and reflects the V-polarized photon. BD--beam displacer, which can couple H- and V-polarized components from different spatial modes. HWP--half-wave plate set at $45^\circ$.}
\label{fig1}
\end{figure}

Here we propose and report the first high-efficiency long-distance polarization entanglement purification using only one pair of hyperentanglement. We also demonstrate its practical application in entanglement-based QKD~\cite{E91}. We show that the EPP using two copies~\cite{purification1} and subsequent experiments~\cite{experiment1,experiment2,experiment3,experiment4} is not necessary and polarization entanglement can be purified using entanglement in other DOF. Moreover, the double-pair emission noise using spontaneous parametric down-conversion (SPDC) source is removed automatically and the purification efficiency can be greatly increased in a second time. A general protocol is shown in Fig.~\ref{fig1}a. In our experiment, we use hyperentanglement encoded on polarization and spatial modes. As shown in Fig.~\ref{fig1}b, a hyperentangled state $|\phi\rangle=|\Phi^{+}\rangle_{ab}\otimes|\phi^{+}\rangle_{ab}$ is distributed to Alice and Bob. $|\Phi^{+}\rangle_{ab}$ is one of the polarization Bell states
$ |\Phi^{\pm}\rangle_{ab}=\frac{1}{\sqrt{2}}(|H\rangle_{a}|H\rangle_{b}\pm|V\rangle_{a}|V\rangle_{b})$,
and $|\Psi^{\pm}\rangle_{ab}=\frac{1}{\sqrt{2}}(|H\rangle_{a}|V\rangle_{b}\pm|V\rangle_{a}|H\rangle_{b})$.  $|\phi^{+}\rangle_{ab}$ is one of the spatial mode Bell states $|\phi^{\pm}\rangle_{ab}=\frac{1}{\sqrt{2}}(|a_{1}\rangle|b_{1}\rangle\pm|a_{2}\rangle|b_{2}\rangle)$,
and $|\psi^{\pm}\rangle_{ab}=\frac{1}{\sqrt{2}}(|a_{1}\rangle|b_{2}\rangle\pm|a_{2}\rangle|b_{1}\rangle)$, where $H (V)$ denotes horizontal (vertical) polarization, and $a_{1}$, $b_{1}$, $a_{2}$, and $b_{2}$ are the spatial modes. The noise channel makes the hyperentangled state become a mixed state as  $\rho_{ab}=\rho^{P}_{ab}\otimes\rho^{S}_{ab}$ with
\begin{eqnarray}
  \rho^{P}_{ab}=F_{1}|\Phi^{+}\rangle_{ab}\langle\Phi^{+}|+(1-F_{1})|\Psi^{+}\rangle_{ab}\langle\Psi^{+}|,\label{mixed1}
   \end{eqnarray}
  and
    \begin{eqnarray}
  \rho^{S}_{ab}=F_{2}|\phi^{+}\rangle_{ab}\langle\phi^{+}|+(1-F_{2})|\psi^{+}\rangle_{ab}\langle\psi^{+}|.\label{mixed2}
\end{eqnarray}
The principle of purification is to select the cases in which the photons are in the output modes D$_{1}$D$_{2}$ or D$_{3}$D$_{4}$~\cite{Supplementary} and we can obtain a new mixed state
\begin{eqnarray}
  \rho'_{ab}=F'|\Phi^{+}\rangle_{ab}\langle\Phi^{+}|+(1-F')|\Psi^{+}\rangle_{ab}\langle\Psi^{+}|\label{mixed3}.
\end{eqnarray}
Here $F'=\frac{F_{1}F_{2}}{F_{1}F_{2}+(1-F_{1})(1-F_{2})}$. If $F_{1}> \frac{1}{2}$ and $F_{2}> \frac{1}{2}$, we can obtain $F'> F_{1}$ and  $F'> F_{2}$.

\begin{figure*}[tbph]
\begin{center}
\includegraphics [width= 2\columnwidth]{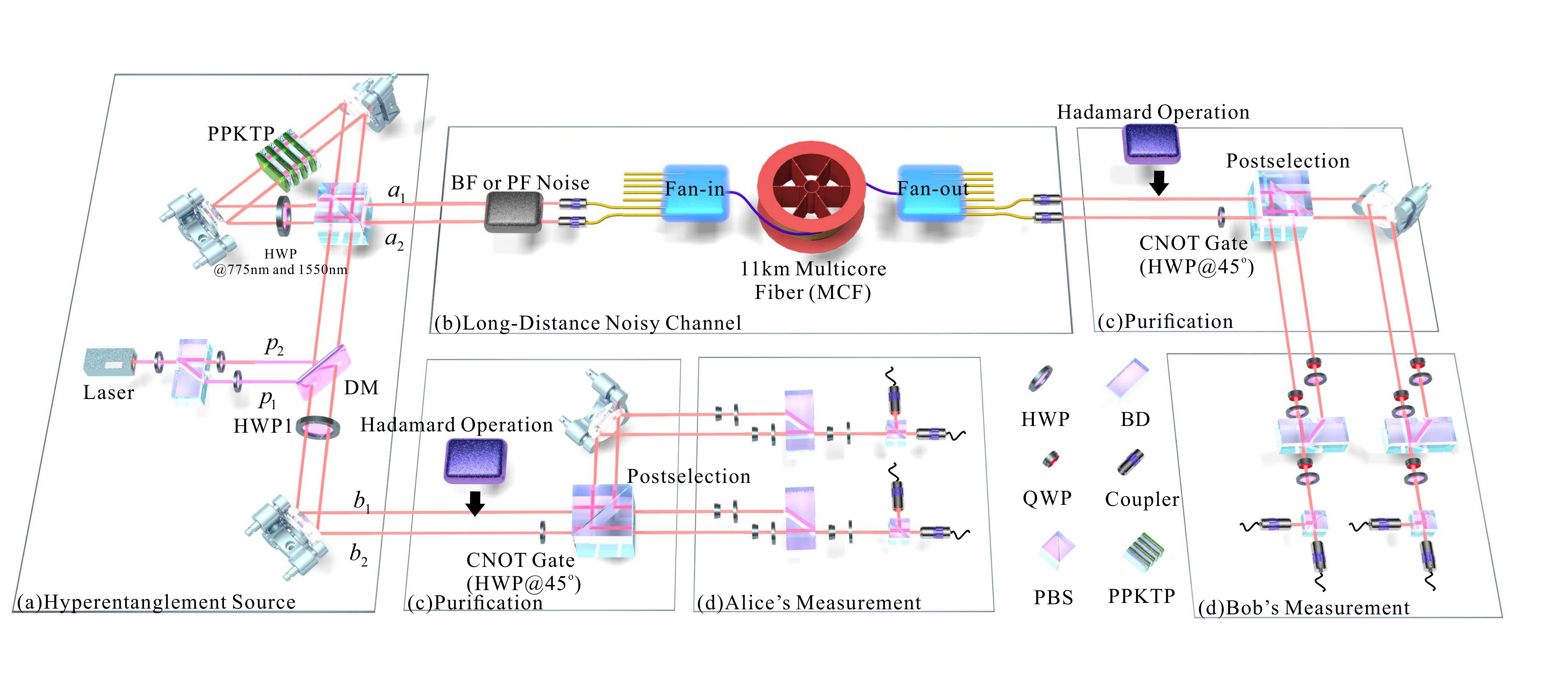}
\end{center}
\caption{\textbf{Experimental Setup.} \textbf{a}, Preparation of hyperentangled state.  The pump light beam is separated into two spatial modes by the BD. These two beams are injected into a Sagnac interferometer to pump a type-II cut periodically poled potassium titanyl phosphate (PPKTP) crystal (1mm$\times$7mm$\times$10mm) and generate the two-photon polarization entanglement $1/\sqrt{2}(|HV\rangle+|VH\rangle)$ in each spatial mode. After HWP1 (set at $45^\circ$), the hyperentangled state $|\Phi^{+}\rangle_{ab}\otimes|\phi^{+}\rangle_{ab}=1/\sqrt{2}(|HH\rangle+|VV\rangle)\otimes 1/\sqrt{2}(|a_{1}b_{1}\rangle+|a_{2}b_{2}\rangle)$ is generated. \textbf{b}, Quantum noisy channel. This channel is divided into two parts. The first part is a controllable spatial mode and polarization bit flip (BF) or phase flip (PF) loading noise setup. The other part is an 11~km multicore fiber (MCF). \textbf{c}, Purification process. This process is also divided into two steps. The first step is a controlled-NOT (CNOT) gate between the spatial mode and polarization DOF. The setup consists of an HWP (set at $45^{\circ}$) placed on spatial modes $a_{2}$ and $b_{2}$. The PBS is used for post-selection of the polarization qubit, and the spatial mode states of the same polarization are preserved while different polarization states are ignored. Hadamard operations are needed to convert the PF noise to BF noise before purification. \textbf{d}, Quantum state tomography setup. BD--beam displacer, DM--dichroic mirror, HWP--half-wave plate, QWP--quarter-wave plate, PBS--polarizing beam splitter. }
\label{fig2}
\end{figure*}

\begin{figure*}[tbph]
\begin{center}
\includegraphics [width= 2\columnwidth]{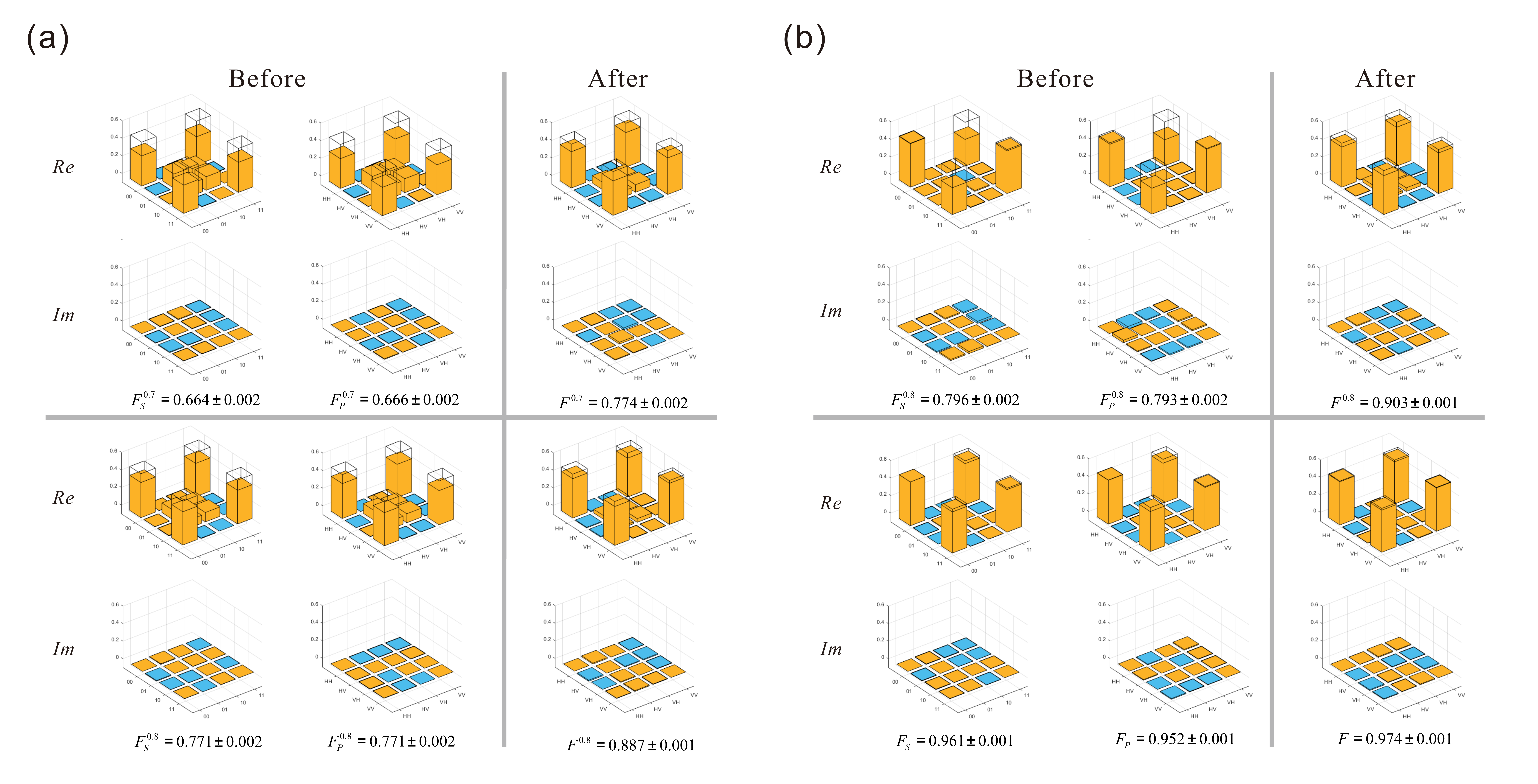}
\end{center}
\caption{\textbf{Experimental results.} \textbf{a}, Results before and after BF noise purification. On the left are the density matrices of polarization and path-entangled states before purification, and on the right are the density matrices of path states after purification. The yellow column represents a value greater than 0, and the blue column represents a value less than 0. The transparent column represents the value of the ideal maximally entangled state. The fidelity of the purified quantum state has obviously been improved significantly. \textbf{b}, Results before and after PF noise purification. The first half is the result of loading $20\%$ PF noise. The second half is the result of noise introduced only by the MCF.}
\label{fig3}
\end{figure*}

To demonstrate the purification, we first generated one pair of hyperentangled state. As shown in Fig.~\ref{fig2}a, a continuous-wave (CW) laser operated at 775~nm is separated into two spatial modes ($p_{1}$ and $p_{2}$) by a beamdisplacer and then injected to a polarization Sagnac interferometer to generate polarization-entangled photon pair~\cite{Kim2006} in each spatial modes~\cite{highdimension2,highdimension7,Guo2018}. Noticed that we use a CW laser, the final state is the superposition of the states in each mode. Thus we can generate the hyperentanglement $|\phi\rangle=|\Phi^{+}\rangle_{ab}\otimes|\phi^{+}\rangle_{ab}$ by tune the relative phase between the two spatial modes. We used 200~mW pumped light to excite 2400 photon pairs/s. The coincidence efficiency of the entangled source is $18\%$. To show the performance of entanglement purification in the noisy channel, we distributed the hyperentangled state over an 11~km multicore fiber (MCF)~\cite{highdimension7,Xavier2020,Canas2017,Bacco2019}. The difficulty of long-distance distribution of polarization and spatial mode hyperentanglement is maintaining the coherence and phase stability between different paths. The MCF provides an ideal platform for distributing spatial-mode states over a long distance. The distance between the nearest two cores of the MCF is very small (approximate 41.5~$\mu$m), and the noises of different paths are very close, so it can maintain coherence~\cite{highdimension7,Xavier2020,Canas2017,Bacco2019}. However, there still have some other difficulties, such as the polarization-maintaining and group delay mismatch. To overcome these obstacles, we used a phase-locking system to ensure the effective distribution of hyperentanglement~\cite{highdimension7,Supplementary}. In Fig.~\ref{fig2}b, the hyperentangled state $|\phi\rangle$ was distributed over 11~km in the MCF. During distribution, a small bit flip (BF) error ($|\Psi^{+}\rangle_{ab}$ and $|\psi^{+}\rangle_{ab}$) and small phase flip (PF) error ($|\Phi^{-}\rangle_{ab}$ and $|\phi^{-}\rangle_{ab}$) were introduced by the fiber noise environment. The fidelities of the hyperentangled state in the polarization and spatial modes were $F_{P}=0.961\pm0.001$ and $F_{S}=0.952\pm0.001$, respectively. Here, we use superconducting single photon detectors to detect each photon, the efficiency is 80\%, and the dark count rate is approximate 300~Hz. Including all the losses, the coincidence efficiency was $\sim8.1\%$, and the coincident count rate was 600~Hz.

In our experiment, we added symmetrical BF noise to both the polarization and spatial mode DOFs, so that $F_{P}\approx F_{S}\approx F$. The BF noise loading setup~\cite{Supplementary} can add any proportion of BF noise ($|\Psi^{+}\rangle_{ab}$ and $|\psi^{+}\rangle_{ab}$) to the hyperentangled state of the polarization and spatial modes. We loaded $20\%$ BF noise into the ideal state, and when it was combined with the MCF, the fidelities of the polarization and spatial mode states were $F_{P}^{0.8}=0.771\pm0.002$ and $F_{S}^{0.8}=0.771\pm0.002$, respectively. When $30\%$ BF noise was added, the fidelities of the polarization and spatial mode states were $F_{P}^{0.7}=0.666\pm0.002$ and $F_{S}^{0.7}=0.664\pm0.002$, respectively.

The purification setup is rather simple and only contains a PBS and an HWP (Fig.~\ref{fig2}c). It is essentially the controlled-NOT (CNOT) gate between the polarization and spatial mode DOFs for a single photon. Unlike the CNOT gate between two photons in polarization, such a CNOT gate works in a deterministic way and does not exploit the auxiliary single photon. The control qubit can be regarded as a spatial mode qubit ($|a_{1}\rangle=|0\rangle$,$|a_{2}\rangle=|1\rangle$), and the target qubit can be regarded as the polarization qubit. The CNOT gate makes $|0\rangle|H\rangle\rightarrow|0\rangle|H\rangle$, $|0\rangle|V\rangle\rightarrow|0\rangle|V\rangle$, $|1\rangle|H\rangle\rightarrow|1\rangle|V\rangle$, and $|1\rangle|V\rangle\rightarrow|1\rangle|H\rangle$. After the CNOT operation, the second operation is to postselect the polarization qubit. Through the PBS, the spatial mode states with the same polarization are retained, and different polarization states are ignored. The purification process is completed. The experimental results show that the fidelity of the purified state was significantly improved for BF noise (Fig.~\ref{fig3}a). For $20\%$ BF noise, the fidelity after purification became $F^{0.8}=0.887\pm0.001$, which is very close to the theoretically predicted value $F=0.896$. For $30\%$ BF noise, the fidelity after purification became $F^{0.7}=0.774\pm0.002$, which is also very close to the theoretical value $F=0.778$.

BF and PF noise can be converted to each other through the Hadamard operations~\cite{Supplementary}. We also show that our protocol is still feasible in the case of PF noise. A PF noise proportion of $20\%$ ($|\Phi^{-}\rangle_{ab}$ and $|\phi^{-}\rangle_{ab}$) was loaded into the hyperentangled state. When this was combined with the MCF noise, the fidelities of the polarization and spatial mode states were $F_{P}^{0.8}=0.793\pm0.002$ and $F_{S}^{0.8}=0.796\pm0.002$, respectively. Differently from the case of BF noise, we first converted PF noise into BF noise through Hadamard operations and then completed entanglement purification. The fidelity after purification is $F^{0.8}=0.903\pm0.001$, which is also very close to the theoretical value $F=0.932$. For hyperentangled states with only MCF noise, we found that PF noise ($\sim3.3\%$) was much higher than BF noise ($\sim1.1\%$). After the purification, the fidelity was $F=0.974\pm0.001$. This is  higher than the fidelity of the polarization or spatial mode state before purification, which shows that our purification was efficient in fiber distribution.

\begin{figure*}[tbph]
\begin{center}
\includegraphics [width=2\columnwidth]{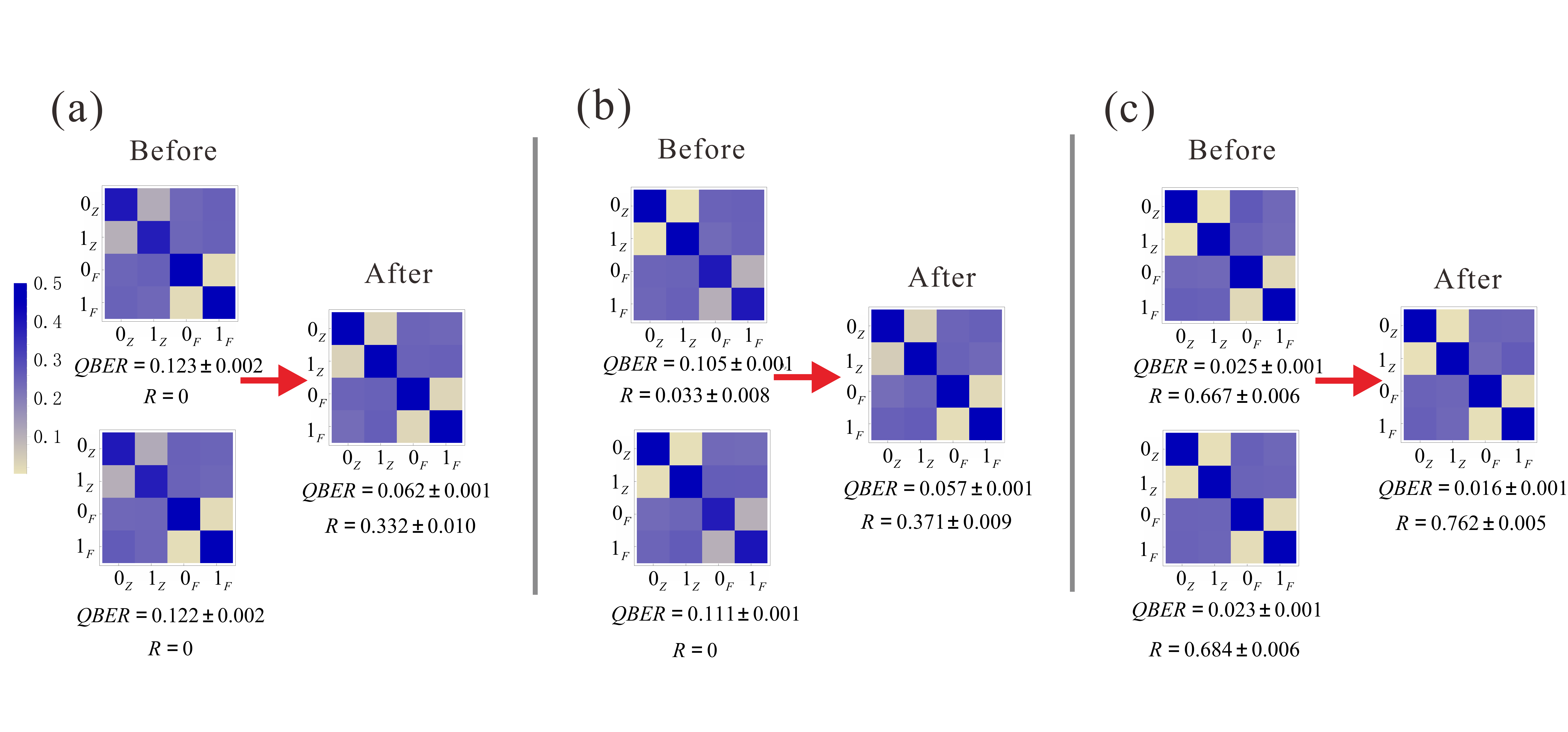}
\end{center}
\caption{\textbf{QKD based on entanglement purification.}
\textbf{a},\textbf{b},\textbf{c} represent the correlations for mutually unbiased bases required for QKD (computational bases ($|0\rangle_{Z},|1\rangle_{Z}$) and Fourier bases $|0\rangle_{F},|1\rangle_{F}$)  before and after purification in the case of loading $20\%$ BF noise, $20\%$ PF noise and using MCF only. The corresponding quantum bit error rate (QBER) and quantum key rate (R) are also presented.
}
\label{fig4}
\end{figure*}

Finally, let us show the practical application of this purification experiment. The first is to increase the secure key rate ~\cite{Shor2000} in entanglement-based QKD~\cite{E91}.  Secure QKD requires that the quantum bit error rate (QBER) is less than $11\%$ (QBER=1-$F$, $F$ is fidelity) \cite{QKD} to generate an effective key rate ($R$). In $20\%$ BF noise and $20\%$ PF noise cases, as shown in Fig.~\ref{fig4}a and Fig.~\ref{fig4}b, after purification, the $R$ increases significantly from $0$ or nearly $0$ to $0.332\pm 0.010$ and $0.371\pm 0.009$. Here $R$ is defined as $R=1-2H(QBER)$, where $H(e)$ represents the Shannon entropy, given as a function of the QBER by $H_{e}\left(e\right)=-\left(1-e\right) \log _{2}\left(1-e\right)-e \log _{2}\left(e\right)$. We also show the improvement of $R$ along a real noise  channel in Fig.~\ref{fig4}c. The second is to distill nonlocality from nonlocal resources~\cite{nonlocal}, which has the potential application to improve the noise tolerance in future device-independent QKD (DI-QKD)~\cite{diqkd}.
Using the reconstructed density matrix, we can calculate the values of Clauser-Horne-Shimony-Holt (CHSH) inequality of these nonlocal quantum states. Initially, for $30\%$ BF noise, $S_{S}=1.837\pm0.006<2$ for spatial mode entanglement and $S_{P}=1.829\pm0.006<2$ for polarization entanglement. After purification, $S=2.128\pm0.006>2$~\cite{Supplementary}.

The integration time of each data point is 60~s, and the count rate of the entangled source after fiber distribution is $\sim600/s$ (before purification). After purification, due to the influence of postselection, the successful events are retained and the failure events are ignored, thus the count rate of the entangled source after purification is reduced respectively. For $30\%$ BF noise, the count rate of purified entangled source is $\sim350/s$, for $20\%$ BF and PF noise, the count rate of purified entangled source is $\sim410/s$.

We propose and demonstrate the first long-distance polarization entanglement purification and show its practical application to increase the secure key rate in entanglement-based QKD and improve the noise tolerance in DI-QKD. Compared with all two-copy EPPs based on  Ref.~\cite{purification1}, our EPP using one pair of hyperentanglement has several advantages. Firstly, this protocol reduces half of the consumption of copies of entangled pairs. Secondly,  benefited from the success probability 100$\%$ CNOT gate between the polarization and spatial inner DOFs, the purification efficiency of this EPP is four times than that of two-copy EPPs in Refs.~\cite{purification2,experiment1,Supplementary}.  Thirdly, if we consider the experimental implementation (SPDC sources), the double-pair emission noise in generating two clean pairs can be removed automatically and the purification efficiency can be estimated as $1.65\times 10^{3}$ times than the EPPs using two pairs entangled states with SPDC sources. The total purification efficiency can be calculated as $4\times 1.65\times 10^{3}=6.6 \times 10^{3}$  than the EPPs using two pairs entangled states with SPDC sources. It is worth noting that since both outcomes of PBSs are used for postselection, we need two sets of measurement setup at both sides of Alice and Bob. However, in the two copy EPPs~\cite{experiment1}, two photons act as triggers, so two additional measurement setups are also needed. Our protocol is general and can be effectively extended to other DOFs of photons, such as the time bin~\cite{Martin2017}, frequency~\cite{Kues2017}, and orbital angular momentum~\cite{multiteleportation}, to perform multi-step purification to improve the fidelity of entanglement further. Moreover, if combining with the suitable high-capacity and high-fidelity quantum memory~\cite{memory2} and entanglement swapping~\cite{multiteleportation,highdimension5}, the approach presented here could be extended to implement the full repeater protocol and large scale quantum networks, enabling a polynomial scaling of the communication rate with distance.

\begin{acknowledgments} This work was supported by the National Key Research and Development Program of China (No.\ 2017YFA0304100, No. 2016YFA0301300 and No. 2016YFA0301700), National Natural Science  Foundation of China (Nos. 11774335, 11734015, 11874345, 11821404, 11904357, 11974189), the Key Research Program of Frontier Sciences, CAS (No.\ QYZDY-SSW-SLH003), Science Foundation of the CAS (ZDRW-XH-2019-1), the Fundamental Research Funds for the Central Universities, Science and Technological Fund of Anhui Province for Outstanding Youth (2008085J02), Anhui Initiative in Quantum Information Technologies (Nos.\ AHY020100, AHY060300).
\end{acknowledgments}

\clearpage
\setcounter{table}{0}
\renewcommand{\thetable}{S\arabic{table}}
\setcounter{figure}{0}
\renewcommand{\thefigure}{S\arabic{figure}}
\setcounter{equation}{0}
\renewcommand{\theequation}{S\arabic{equation}}

\textbf{SUPPLEMENTARY INFORMATION}\\
\noindent\textbf{The protocol of entanglement purification using hyperentanglement}\\
As shown in Fig.~1, the hyperentanglement source (S) generates one pair of hyperentangled state $|\phi\rangle=|\Phi^{+}\rangle_{ab}\otimes|\phi^{+}\rangle_{ab}$. After the photons transmit across the channel, the polarization and spatial mode entanglements become a mixed state $\rho_{ab}=\rho^{P}_{ab}\otimes\rho^{S}_{ab}$.
$|\Psi^{+}\rangle_{ab}$ and $|\psi^{+}\rangle_{ab}$ mean that bit-flip error occurs for both polarization and spatial mode entanglements. Thus, the initial hyperentangled state becomes a probabilistic mixture of four states. These states are: $|\Phi^{+}\rangle_{ab}\otimes|\phi^{+}\rangle_{ab}$ with a probability of $F_{1}F_{2}$, $|\Phi^{+}\rangle_{ab}\otimes|\psi^{+}\rangle_{ab}$ with a probability of $F_{1}(1-F_{2})$, $|\Phi^{+}\rangle_{ab}\otimes|\phi^{+}\rangle_{ab}$ with a probability of $(1-F_{1})F_{2}$, and $|\Psi^{+}\rangle_{ab}\otimes|\psi^{+}\rangle_{ab}$ with a probability of $(1-F_{1})(1-F_{2})$. The first state,  $|\Phi^{+}\rangle_{ab}\otimes|\phi^{+}\rangle_{ab}$, leads the two photons in the output modes D$_{1}$D$_{2}$, namely $\frac{1}{\sqrt{2}}(|H\rangle_{D_{1}}|H\rangle_{D_{2}}+|V\rangle_{D_{1}}|V\rangle_{D_{2}})$ or in the output modes D$_{3}$D$_{4}$, namely  $\frac{1}{\sqrt{2}}(|H\rangle_{D_{3}}|H\rangle_{D_{4}}+|V\rangle_{D_{3}}|V\rangle_{D_{4}})$. The state $|\Psi^{+}\rangle_{ab}\otimes|\psi^{+}\rangle_{ab}$ also leads the two photons in the output mode D$_{1}$D$_{2}$ or D$_{3}$D$_{4}$. They are in the state $\frac{1}{\sqrt{2}}(|H\rangle_{D_{1}}|V\rangle_{D_{2}}+|V\rangle_{D_{1}}|H\rangle_{D_{2}})$ or $\frac{1}{\sqrt{2}}(|H\rangle_{D_{3}}|V\rangle_{D_{4}}+|V\rangle_{D_{3}}|H\rangle_{D_{4}})$. The other two states, $|\Phi^{+}\rangle_{ab}\otimes|\psi^{+}\rangle_{ab}$ and $|\Psi^{+}\rangle_{ab}\otimes|\phi^{+}\rangle_{ab}$, cannot lead the two photons in the output modes D$_{1}$D$_{2}$ or D$_{3}$D$_{4}$. They are in the output modes D$_{1}$D$_{4}$ or D$_{2}$D$_{3}$. Finally, by selecting the cases in which the photons are in the output modes D$_{1}$D$_{2}$ or D$_{3}$D$_{4}$, we can obtain a new mixed state:

\begin{eqnarray}
  \rho'_{ab}=F'|\Phi^{+}\rangle_{ab}\langle\Phi^{+}|+(1-F')|\Psi^{+}\rangle_{ab}\langle\Psi^{+}|\label{mixed3}.
   \end{eqnarray}
Here $F'=\frac{F_{1}F_{2}}{F_{1}F_{2}+(1-F_{1})(1-F_{2})}$. If $F_{1}> \frac{1}{2}$ and $F_{2}> \frac{1}{2}$, we can obtain $F'> F_{1}$ and  $F'> F_{2}$.
In general, an error model  contains not only bit-flip error, but also phase-flip error. The phase-flip error can be converted into the bit-flip error by adding the unitary operations and can also be purified.\\

\noindent\textbf{Multicore fiber (MCF) and optical fiber locking system}

\begin{figure*}[tbph]
\begin{center}
\includegraphics [width= 1.5\columnwidth]{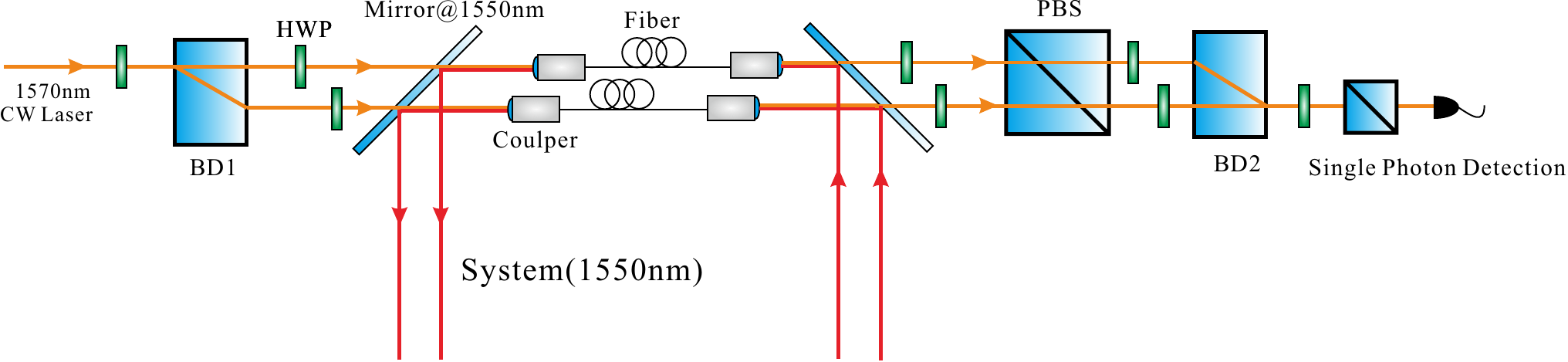}
\vspace{-0.5cm}
\end{center}
\caption{\textbf{Optical fiber phase-locking system.} To reduce the effect of the reference light (orange solid line) on the signal light (red solid line) of the system, they propagate in opposite directions. The light intensities from both are weak. BD1 and BD2 constitute an MZ interferometer used to lock the phase. }
\label{figs1}
\end{figure*}

The MCF we used here was 11~km long and had a core diameter of 8~$\mu$m and seven cores with a core-to-core distance of 41.5~$\mu$m. Each core supported a single optical mode and had transmission characteristics similar to those of a standard single-mode fiber (SMF). The crosstalk between the cores was -45~dB/100~km. The MCF was divided into seven separate SMFs through a fan in and fan out. The average loss (including the insertion loss) of the fiber was 5.13~dB. The technical challenge of the spatial-mode entanglement in long-distance distribution is to maintain phase stability along different spatial modes. In our protocol, the phase between different MCF cores is stable. However, considering the fan in and fan out, the relative phase still needs to be locked. We used a fiber-locking system to lock the relative phase of the two cores. To reduce the disturbance of the quantum state by optical-fiber phase locking, the reference light (1570~nm CW light) was opposite to the system light. The reference light passed through BD1 and was divided into two beams (Fig.~\ref{figs1}), which were coupled to the fan-out fibers using mirrors. The light in the lower path was reflected by the mirror of a piezoelectric ceramic material (PZT), which was used to change the length of the optical path to stabilize the phase. The reference light was then emitted from the other end of the fan in the optical fiber, and two Mach-Zehnder (MZ) interferometers were formed by BD1,BD2. The phase between BDs was stable, and all the phase changes were due to the instability of the two MCF cores. Hence, we only needed to use the PZT to adjust the position of the mirror according to the signal of the MZ interferometer composed of BD1 and BD2 to lock the phase of the optical fiber.\\

\noindent\textbf{BF and PF loading setup}

\begin{figure}[tbph]
\begin{center}
\includegraphics [width= 1\columnwidth]{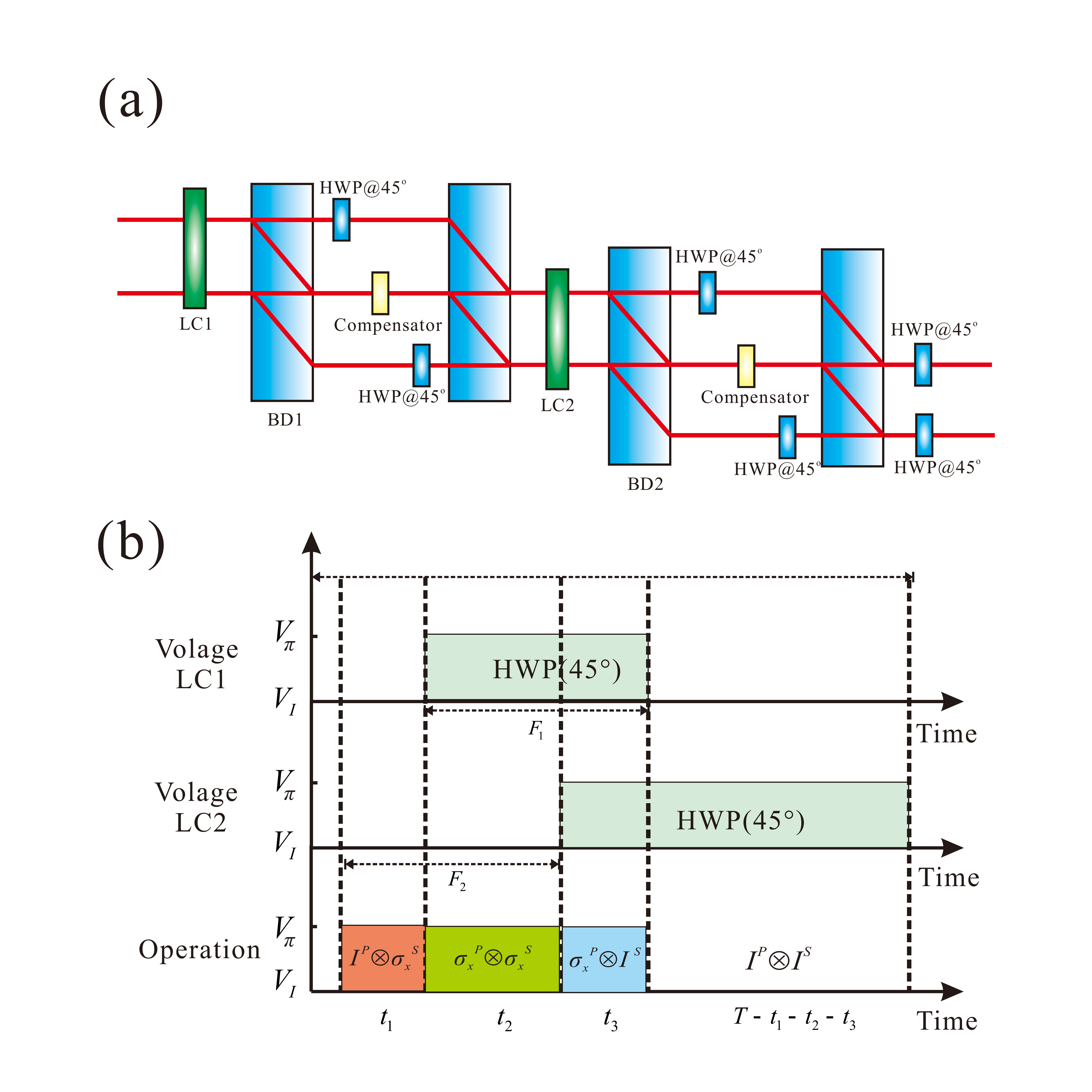}
\vspace{-0.5cm}
\end{center}
\caption{\textbf{Bit-flip noise setup.} \textbf{a}, The setup consists of HWPs, compensations, BDs, LC1($@45^{\circ}$), and LC2($@45^{\circ}$). When the voltage $V_{\pi}$ or $V_{I}$ is applied to the LC1, a $\sigma_{x}^{P}$ or $\sigma_{I}^{P}$ operation is performed on the polarization qubit, respectively. When V$_{I}$ or V$_{\pi}$ is applied to the LC2, a $\sigma_{x}^{S}$ or $\sigma_{I}^{S}$ operation is performed on the spatial mode qubit, respectively. \textbf{b}, The LC activation times, $t_{1}$, $t_{2}$, and $t_{3}$, correspond to the three kinds of BF noise, $\sigma_{I}^{P}\otimes\sigma_{x}^{S} $, $\sigma_{x}^{P}\otimes \sigma_{x}^{S}$, and $\sigma_{x}^{P}\otimes \sigma_{I}^{S}$, respectively. By adjusting $t_{1}$, $t_{2}$, and $t_{3}$, any proportion of BF noise can be loaded.}
\label{figs2}

\end{figure}

\begin{figure}[tbph]
\begin{center}
\includegraphics [width= 1\columnwidth]{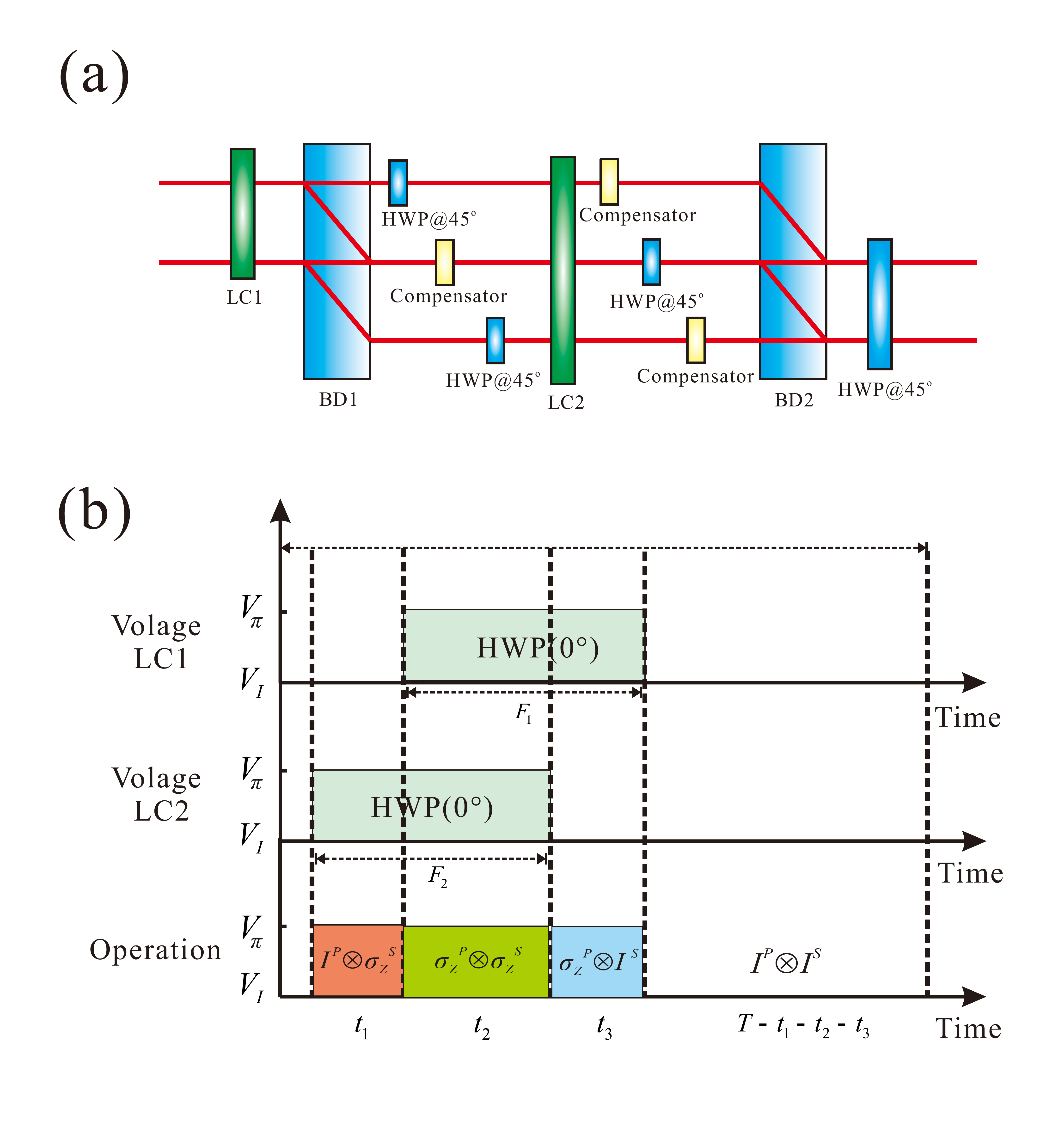}
\vspace{-1.0cm}
\end{center}
\caption{\textbf{Phase flip noise Setup.} \textbf{a}, The setup consists of HWPs, compensations, BDs, LC1($@0^{\circ}$), and LC2($@0^{\circ}$). When the voltage $V_{\pi}$ or $V_{I}$ is applied to the LC1, the operation $\sigma_{z}^{P}$ or $\sigma_{I}^{P}$ is performed on the polarization qubit, respectively. When the voltage $V_{\pi}$ or $V_{I}$ is applied to the LC2, the operation $\sigma_{z}^{S}$ or $\sigma_{I}^{S}$ is performed on the spatial mode qubit, respectively. \textbf{b}, Photon B is transmitted through a phase flip noise channel. This is implemented by liquid crystals LC1 and LC2. The LC activation times, $t_{1}$, $t_{2}$, and $t_{3}$, correspond to the three kinds of phase flip noise, $\sigma_{I}^{P}\otimes\sigma_{z}^{S} $, $\sigma_{z}^{P}\otimes \sigma_{z}^{S}$, and $\sigma_{z}^{P}\otimes \sigma_{I}^{S}$, respectively.  }
\label{figs3}
\end{figure}

In this section, we introduce the BF and PF noise loading setup in detail. These setups are composed of HWPs, liquid crystals (LCs), and PBSs and BDs, which can achieve any proportion of BF (Fig.~\ref{figs2}) or PF (Fig.~\ref{figs3}) noise loading. When voltage V$_{\pi}$ or V$_{I}$ is applied to LC1 in Fig.~\ref{figs2}(a), the operation $\sigma^{P}_{x}\equiv|H\rangle\langle V|+|V\rangle\langle H|$ or $\sigma^{P}_{I}\equiv|H\rangle\langle H|+|V\rangle\langle V|$ is applied to the polarization qubit, respectively. Meanwhile, when voltage V$_{I}$ or V$_{\pi}$ is applied to LC2, the operation $\sigma^{S}_{x}\equiv|b_{1}\rangle\langle b_{2}|+|b_{2}\rangle\langle b_{1}|$ or $\sigma^{S}_{I}\equiv|b_{1}\rangle\langle b_{1}|+|b_{2}\rangle\langle b_{2}|$ is applied to the spatial mode qubit, respectively. We adjust the temporal delay between the intervals corresponding to V$_{I}$ and V$_{\pi}$ applied to the two LCs. In Fig.~\ref{figs2}(b), we define $T$ as the period of the LC activation cycle, and $t_{1}$, $t_{2}$, $t_{3}$, and $T-t_{1}-t_{2}-t_{3}$ as the activation times of the operations $\sigma^{S}_{x}\otimes \sigma^{P}_{I}$, $\sigma^{S}_{x}\otimes \sigma^{P}_{x}$, $\sigma^{S}_{I}\otimes \sigma^{P}_{x}$, and $\sigma^{S}_{I}\otimes \sigma^{P}_{I}$. In the interval between $t_{2}$ and $t_{3}$, V$_{\pi}$ is added to LC1 to realize the operation $\sigma^{P}_{x}$ of the polarization qubit. In the interval between $t_{1}$ and $t_{2}$, V$_{I}$ is applied to LC2 to realize $\sigma^{S}_{x}$ for the spatial mode qubit. In this way, we can change the noise ratio by adjusting the interval times $t_{1}$, $t_{2}$, and $t_{3}$ for a fixed period $T$. In the actual experiment, we take a period $T=15 s$.

The PF noise loading setup is similar to that of BF noise, as shown in Fig.~\ref{figs3}(a). When V$_{\pi}$ or V$_{I}$ is applied in LC1, $\sigma^{P}_{z}\equiv|H\rangle\langle H|-|V\rangle\langle V|$ or $\sigma^{P}_{I}\equiv|H\rangle\langle H|+|V\rangle\langle V|$ is added to the polarization qubits, respectively. When V$_{\pi}$ or V$_{I}$ is applied to LC2, $\sigma^{S}_{z}\equiv|b_{1}\rangle\langle b_{1}|-|b_{2}\rangle\langle b_{2}|$ or $\sigma^{S}_{I}\equiv|b_{1}\rangle\langle b_{1}|+|b_{2}\rangle\langle b_{2}|$ is added on the spatial mode qubit, respectively. We can use the same loading time sequence as for BF noise, as shown in Fig.~\ref{figs3}(b). This way, we can load any proportion of PF error on the qubit in both the polarization and spatial mode DOFs.\\

\noindent\textbf{Hadamard operation for BF noise and PF noise conversion}

\begin{figure}[tbph]
\vspace{0.5cm}
\begin{center}
\includegraphics [width= 1\columnwidth]{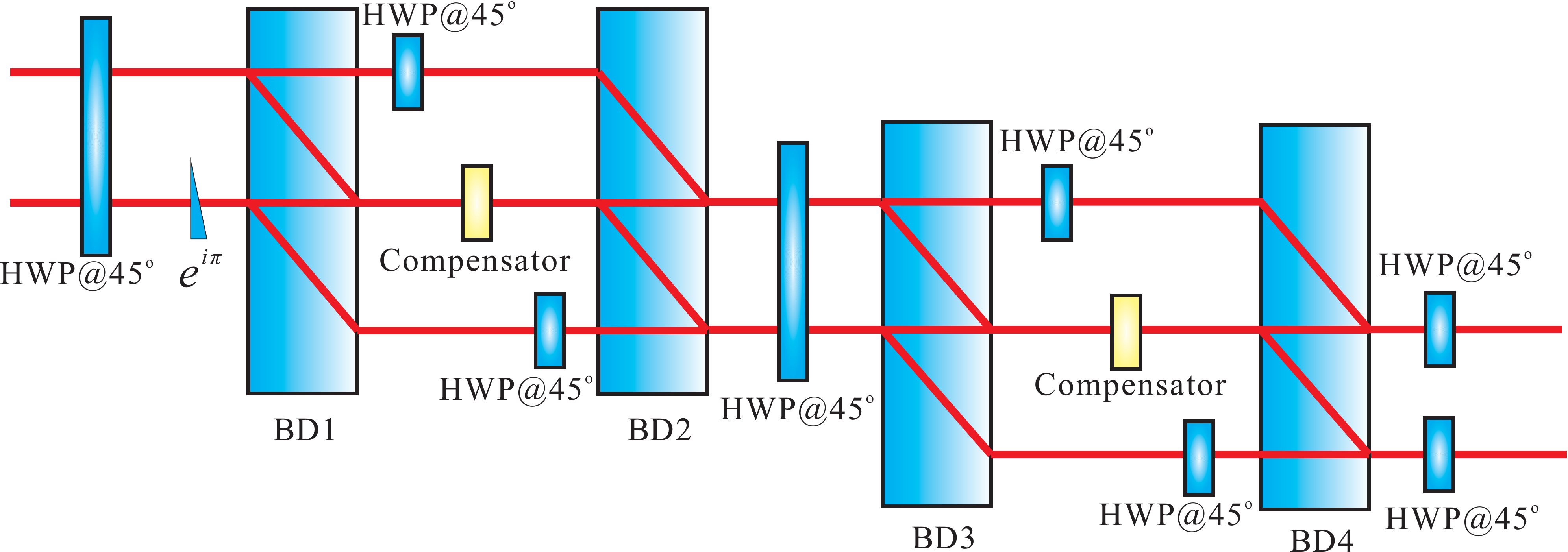}
\end{center}
\caption{\textbf{Setup for Hadamard operation.} The Hadamard operation on the polarization entanglement and spatial mode entanglement makes $|\phi^{+}\rangle\leftrightarrow|\phi^{+}\rangle$, $|\phi^{-}\rangle\leftrightarrow|\psi^{+}\rangle$, $|\psi^{+}\rangle\leftrightarrow|\phi^{-}\rangle$, and $|\psi^{-}\rangle\leftrightarrow|\psi^{-}\rangle$, respectively. The setup for the Hadamard operation for both polarization and spatial mode qubits is similar to the setup for BF noise. Here, we replace LCs with HWPs ($@22.5^{\circ}$). The fidelity of this Hadamard operation is $F\sim0.997$. }
\label{figs4}
\end{figure}

The single qubit unitary operation is important in entanglement purification. If some unitary operations (such as Hadamard operation) are performed where both Alice and Bob are, BF noise and PF noise can be transformed into each other ($|\phi^{+}\rangle\leftrightarrow|\phi^{+}\rangle$, $|\phi^{-}\rangle\leftrightarrow|\psi^{+}\rangle$, $|\psi^{+}\rangle\leftrightarrow|\phi^{-}\rangle$, and $|\psi^{-}\rangle\leftrightarrow|\psi^{-}\rangle$). As shown in Fig.~\ref{figs4}, the experimental setup for the Hadamard operation of the polarization and spatial modes includes HWPs and BDs. Through the setup, the Hadamard operation for polarization and path operation can be expressed as:
\begin{equation}
H_{P}\otimes H_{S}=\left(
\begin{array}{cc}
1 & 1 \\
1 & -1 \\
\end{array}
\right)\otimes
\left(
\begin{array}{ccc}
1 & 1 \\
1 & -1\\
\end{array}
\right).
\end{equation}\\

\noindent\textbf{Polarization-spatial mode hybrid tomography}

 \begin{figure}[tbph]
\begin{center}
\includegraphics [width= 1\columnwidth]{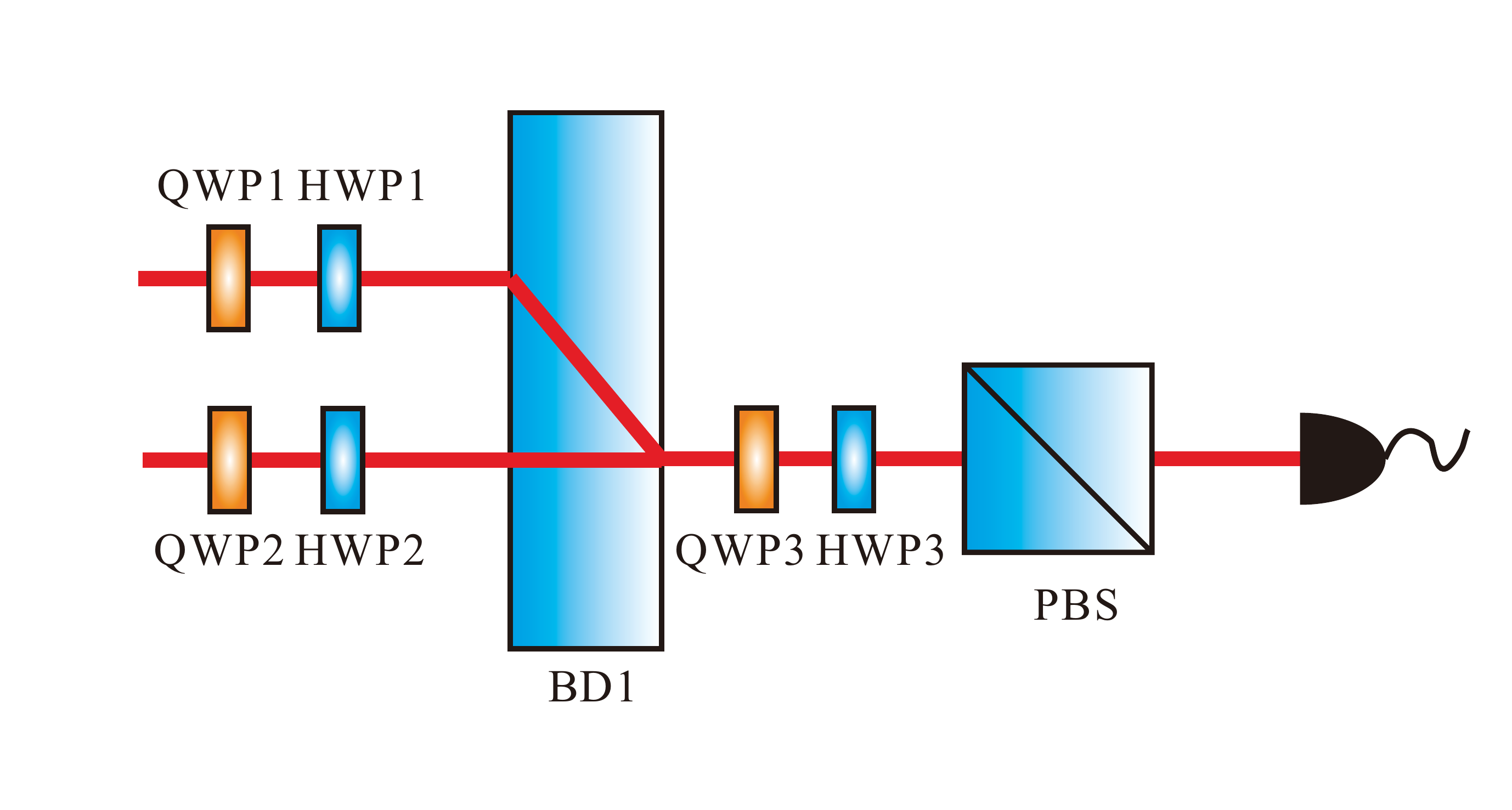}
\vspace{-0.5cm}
\end{center}
\caption{\textbf{ Measurement setup of polarization and spatial mode qubit.} }
\label{figs5}
\end{figure}

As shown in Fig. \ref{figs5}, we use a BD, HWPs, QWPs, and a PBS to perform tomographic analysis of polarization and spatial mode states. When we need to measure one of the DOFs (polarization or spatial mode), we need to trace the other DOF and then conduct standard qubit tomography. If we need to do tomography for the spatial mode DOF, we set HWP1-2 and QWP1-2 appropriately so that the upper and lower $|H\rangle$ polarization states (or $|V\rangle$ polarization states) pass the BD. Then we use HWP3, QWP3, and PBS to complete the tomography for the path qubit. If we need to measure the polarization qubit state, we first select one of the paths $|0\rangle$ (or $|1\rangle$) through HWP3, QWP3, and PBS and then complete the polarization state of the qubit state through HWP1-2, QWP1-2, and BD.  \\

\noindent\textbf{Fidelity estimation}

We can estimate the fidelity of the state after purification according to the density matrix of the state before purification. We take the purification process of loading $20\%$ BF noise as an example. During distribution, small amounts of BF and PF noise are introduced by the fiber noise environment. The fidelities of the hyperentangled state in the polarization and spatial modes are $F_{P}=0.965\pm0.001$ and $F_{S}=0.953\pm0.001$ after distribution. After loading $20\%$ BF noise, the density matrix of the polarization state $\rho^{0.8}_{P}$ and spatial mode state $\rho^{0.8}_{S}$ becomes

\begin{widetext}
\begin{eqnarray}
\rho^{0.8}_{P}=\left(
\begin{array}{cccc}
0.397 + 0.000i & -0.001 - 0.003i &  0.004 - 0.005i &  0.377 - 0.003i \\
-0.001 + 0.003i &  0.111 + 0.000i &  0.090 + 0.001i & -0.006 + 0.004i \\
0.004 + 0.005i &  0.090 - 0.001i &  0.102 + 0.000i & -0.003 + 0.001i \\
0.377 + 0.003i & -0.006 - 0.004i & -0.003 - 0.001i &  0.390 + 0.000i \\
\end{array}
\right), \\
\rho^{0.8}_{S}=\left(
\begin{array}{cccc}
0.400 + 0.000i &  0.002 + 0.008i & 0.005 + 0.001i &  0.379 - 0.002i \\
0.002 - 0.008i &  0.109 + 0.000i  & 0.092 + 0.003i & -0.006 + 0.001i \\
0.005 - 0.001i &  0.092 - 0.003i &  0.107 + 0.000i & -0.000 - 0.002i \\
0.377 + 0.002i & -0.006 - 0.001i & -0.000 + 0.002i &  0.384 + 0.000i \\
\end{array}
\right).
\end{eqnarray}
\end{widetext}

We then do the CNOT operation in polarization and spatial DOF at Alice and Bob respectively. After that, the spatial mode states of the same polarization are preserved, and the different polarization spatial mode states are ignored:
\begin{eqnarray}
\rho_{P}=Tr_{(HH\ and\ VV)}(CNOT\rho^{0.8}_{P}\otimes\rho^{0.8}_{S}).
\end{eqnarray}

The estimated theoretical fidelity after purification is $F=0.896$, which is very close to the fidelity of our experiment, $F^{0.8}=0.887$. A similar method can be used to estimate the fidelity after purification under other noise conditions. In the cases of $30\%$BF noise, $20\%$PF noise, and an MCF only, the estimated fidelities after purification are 0.778, 0.932, and 0.985, respectively.\\

\noindent\textbf{Purification efficiency}

We analyse the efficiency of this protocol in detail. The efficiency consists of three parts. The first one comes from the protocol itself. The success probability of the protocol is $P_{P}=F_{1}F_{2}+(1-F_{1})(1-F_{2})$. The second part comes from the contribution of the entanglement sources, i.e., the generation efficiency of the SPDC implementation.  We denote this part of the efficiency as $P_{s}$. The third part comes from transmission losses. In this protocol, the entanglement is distributed to distant locations. To perform the purification, we should ensure the entangled state does not experience loss. Therefore, the transmission efficiency in the optical fiber is $\eta=e^{\frac{-\alpha L}{10}}$ with $\alpha\simeq0.2 db/km$ for 1550~nm \cite{optical}, where $L$ is the entanglement distribution distance. The total efficiency $P_{one}$ can be written as
 \begin{eqnarray}
P_{one}&=&P_{P} \ast P_{s}\ast\eta\nonumber\\
&=&(F_{1}F_{2}+(1-F_{1})(1-F_{2}))\ast P_{s}\ast e^{\frac{-\alpha L}{10}}.
 \end{eqnarray}

We can also estimate the purification efficiency using two pairs of mixed states with linear optics.
In linear optics, the CNOT gate with a success probability of 1/4. Each purification works in a heralded way, and at least one pair of mixed states should be measured. This way, the success probability of the protocol is $P'_{P}=\frac{1}{4}(F_{1}F_{2}+(1-F_{1})(1-F_{2}))$.

For the efficiency of SPDC implementation, the advantages of our protocol and two copy EPPs need to be compared under the same conditions. The ultrafast pulsed laser is usually used in the multiphoton experiments~\cite{experiment1}, while CW laser is used in our experiments. In principle, after strict compensation, our experimental setup can also be pumped by ultrafast pulse laser~\cite{Hu2019}. For comparison with the two copy SPDC implementation, we assume that our photon source is also pumped by ultrafast pulse laser (76M Hz). The success probability of generating two pairs of entangled states is $P^{2}_{s}$. Both pairs experience the noise during the entanglement distribution, and the success probability of obtaining two pairs of entangled states is $\eta^{2}$. Finally, we can estimate the efficiency as

 \begin{eqnarray}
P_{two}&=&\frac{1}{4}P_{P} \ast P^{2}_{s}\ast \eta^{2}\nonumber\\
&=&\frac{1}{4}(F_{1}F_{2}+(1-F_{1})(1-F_{2}))\ast P^{2}_{s}\ast (e^{\frac{-\alpha L}{10}})^{2} .\nonumber\\
 \end{eqnarray}
The efficiency ratio of the two protocols can be written as
\begin{eqnarray}
 \frac{P_{one}}{P_{two}}=\frac{4}{P_{s}\eta}.
  \end{eqnarray}

In the current entanglement source, $P_{s}=C/(\varepsilon^2\times Rep)=2.4\times10^{3}/(0.18^{2}\times76\times10^{6})\sim0.001$ ($C$ is the coincidence count per second before fiber, $\varepsilon$ is the photon coupling efficiency, and $Rep$ is the repetition rate of the pump light). For an 11~km fiber, we can estimate $\eta\sim 0.602$. We obtain $ \frac{P_{one}}{P_{two}}\sim 6.6\times10^{3}$. \\

\noindent\textbf{Purify the nonlocal quantum states from the local quantum states}

Quantum nonlocality is an important feature of entangled states. Many important quantum information tasks rely on quantum nonlocality, such as device independent quantum key distribution \cite{DIQKD}, device independent quantum secure direct communication \cite{DIQSDC}, quantum communication complexity \cite{Brukner2004}. However, not all entangled states can show quantum nonlocality \cite{Horodecki2009}. In our experiment, we also show the ability of distilling nonlocality from local resources. We use Clauser-Horne-Shimony-Holt (CHSH) inequality to verify the nonlocality of quantum states. Alice and Bob measure observables $A_{1}$, $A_{2}$ and $B_{1}$, $B_{2}$ respectively. For local hidden variable models, they are in accordance with:
\begin{eqnarray}
S=\langle A_{1}B_{1}\rangle+\langle A_{1}B_{2}\rangle+\langle A_{2}B_{1}\rangle-\langle A_{2}B_{2}\rangle\leq 2. \end{eqnarray}
When $S>2$, it cannot be explained by the theory of local hidden variables, showing the quantum nonlocality. For a maximally entangled state, the maximal value is $S=2\sqrt{2}$. In our experiment, when PF noise is $30\%$, we can calculate the violation of CHSH inequality before and after purification by the density matrix. Before purification, $S_{S}=1.837\pm0.006 < 2$ for spatial entanglement, $S_{P}=1.829\pm0.006 < 2$ for polarization entanglement. After purification, $S=2.128\pm0.006>2$. This proves that we distill nonlocality from local quantum states. \\


\end{document}